\documentclass{optica-article}

\journal{opticajournal}

\articletype{Research Article}

\usepackage{lineno}

\usepackage{blindtext}
\usepackage{fancyhdr}

\usepackage{graphicx}
\usepackage{dcolumn}
\usepackage{bm}

\usepackage{siunitx}
\usepackage{physics}
\usepackage{float}
\usepackage[normalem]{ulem}

\begin{document}

\pagestyle{fancy}
\fancyhf{}
\cfoot{\thepage}

\title{Monitoring and active stabilization of laser injection locking using beam ellipticity}

\author{Umang Mishra,\authormark{1} Vyacheslav Li,\authormark{1} Sebastian Wald,\authormark{1} Sofia Agafonova,\authormark{1} Fritz Diorico,\authormark{1,*} and Onur Hosten\authormark{1,*}}

\address{\authormark{1}Institute of Science and Technology Austria, Klosterneuburg, Austria}
\email{\authormark{*}fritz.diorico@ist.ac.at} 
\bigskip
\hrule

\begin{abstract*}
We unveil a powerful method for stabilization of laser injection locking based on sensing variations in the output beam ellipticity of an optically seeded laser. The effect arises due to an interference between the seeding beam and the injected laser output. We demonstrate the method for a commercial semiconductor laser without the need for any internal changes to the readily-operational injection locked laser system that was utilized. The method can also be used to increase the mode-hop free tuning range of lasers, and has the potential to fill a void in the low-noise laser industry.\\
\end{abstract*}

\hrule
\bigskip

Semiconductor lasers have been at the forefront of developments in atomic, molecular, and optical physics. They have also found a wide range of applications in the photonics industry. There is a growing need for widely tunable and spectrally pure lasers for next generation optical telecommunications, remote sensing, lidar, emerging quantum technologies and academic research \cite{Blumenthal2020, Bai2022, Ding2022}. These applications often require a narrow spectral linewidth. One of the primary techniques used to narrow the spectral linewidth of semiconductor lasers is injection locking \cite{Buczek1973,Bondiou2000, Wang2014} -- a phenomenon that has been thoroughly studied over the years \cite{Kobayashi1981, Petermann1995,Samutpraphoot2014, Liu2020}. In an externally injection locked laser system, a broader linewidth laser is made to copy the superior spectral properties of a narrower linewidth one through optical seeding of the former by the latter \cite{Nishizawa1975, Lang1982}. Similarly, a laser can be seeded with a spectrally filtered version of its own output to dramatically improve its emission linewidth --  a technique known as self-injection locking \cite{Laurent1989,Dahmani1987,Liang2010}. This is the core principle used in modern lasers with exceptional performance that utilize optical feedback: distributed feedback lasers (DFB), distributed Bragg lasers (DBR), and external cavity lasers (ECL).

The injection locking process in semiconductor lasers depends sensitively on laser current and temperature \cite{Klotzin2014}, yielding a finite frequency range for successful locking. In relation to the injection locking status, the continuous frequency tunability of an optical-feedback based laser is limited to its so-called mode-hop-free tuning range \cite{Gray1991}. Many techniques have been developed for avoiding mode-hops to obtain wide tunability \cite{Schremer1990, Hult2005, Repasky2006, Kasai2012, Gong2014,Chiow2007}. These rely on either an off-line calibration together with feedforward control, or a method of actively monitoring the injection locking status of the laser together with feedback control. 

Active monitoring of the injection locking status is typically essential to ensuring optimal performance. Existing techniques rely on utilizing laser output polarization \cite{Mueller1998} or its noise \cite{Niederriter2021}, laser modulation and harmonic analysis of its spectral response \cite{Ohshima1992,Labaziewicz2007}, tracking the frequency spectrum \cite{Saxberg2016,Ratkoceri2021} or the frequency noise \cite{Chiow2007} of the injected laser, monitoring the output intensity \cite{Chen2021}, 
indirect utilization of RF modulators \cite{Hayasaka2011}, and lastly, utilization of the beam pointing of the seeding laser \cite{Ottaway2001}. Each technique has a different limitation, for example, due to amplitude fluctuations or misalignment drifts. They typically require additional optical instrumentation around the laser system -- like a Fabry-Perot interferometer (FPI), additional optical elements, or active RF modulators. A universal, modulation-free and easily-scalable approach to generating an error signal for injection locking stabilization is key to developing widely tunable laser systems with reliable spectral noise properties. If accomplished without introducing any additional complications to an existing laser system, this can fill a void in the laser industry.

Here, we present a novel method to monitor the injection locking status of a semiconductor laser without making a single change to an already-operational injection-locked laser system. By means of monitoring the beam shape ellipticity of the optically injected laser, we generate a high-quality error signal and stabilize the laser injection locking process. The enabling technique is adapted from the recently developed `squash locking' method \cite{SquashLockArxiv,SquashLockPatent} for locking the frequency of a laser to an optical cavity -- which already demonstrated extremely competitive locking stabilities \cite{SquashLockArxiv}. By means of interference, squash locking utilizes 2$^{nd}$ order Hermite-Gauss spatial modes as a reference to probe the phase of a fundamental Gaussian mode resonating inside of an optical cavity -- in this case, the semiconductor laser cavity. The interference pattern comes as a beam with varying spatial ellipticity as the phase of the fundamental mode changes as a function of its frequency detuning from the cavity. 

For this work, we take advantage of the inherent residual astigmatic beam profile of a semiconductor laser and the fact that the injecting beam is spatially circular. Based on this, we will now describe the basic principle of operation for error signal generation. From the viewpoint of the laser's native modes (astigmatic Hermite-Gauss modes), the incoming circular injecting beam can be viewed as a superposition of mainly the astigmatic fundamental lasing mode and a small amount of 2$^{nd}$ order astigmatic Hermite-Gauss mode (see \cite{SquashLockArxiv} for more detailed mode shape discussions). Whereas the fundamental component injects the laser and resonates inside the lasing cavity, the 2$^{nd}$ order component, to a sufficient approximation, is promptly reflected from the laser output facet without building-up inside the cavity. Thus, irrespective of the injection frequency, the phase of the 2$^{nd}$ order reference mode can be considered constant. However, the phase of the fundamental mode that builds-up and leaves the cavity will clearly have a dependence on the difference between  the cavity resonance frequency and the injecting beam frequency. The interference between these two components manifests as a varying beam ellipticity as either the injecting laser frequency or the effective laser cavity length is scanned. An error signal corresponding to the varying ellipticity can be obtained simply from a quadrant photodiode (QPD) by subtracting the sums of diagonally positioned quadrants.

This squash-locking-inspired injection status monitoring scheme is analog, modulation-free, and insensitive to intensity and misalignment drifts. It yields an error signal with a zero-crossing, which can be used to stabilize the injection locking process through, e.g., a feedback on the injected laser's current. The simplicity and the effectiveness of the method holds potential for semiconductor laser systems with uninterrupted frequency tuning over the laser's entire operating range with uncompromised spectral noise. 

\begin{figure}[!tb]
    \centering
    \includegraphics[width=8cm]{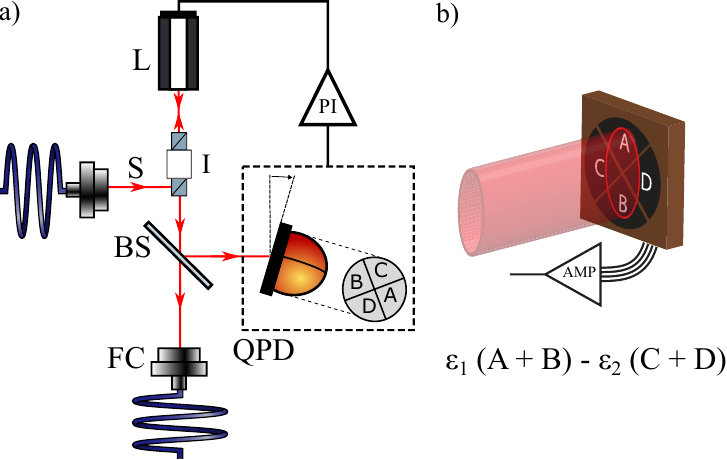}
    \caption{a) Basic scheme to measure and feedback control the injection locking status of an externally `injected' laser. L: the semiconductor laser to be injected; S: optical seeding beam originating from a separate laser; I: optical isolator with an injection port to couple the external seeding laser; BS: beam sampler to sample the laser output shape; QPD: quadrant photodiode to obtain the error signal; FC: fiber collimation package. b) Schematic of the QPD operation. The diagonal channel of the QPD is read out using a transimpedance amplifier circuit which enables the asymmetric detection by providing adjustment for $\varepsilon_{1}$ and $\varepsilon_{2}$. }
    \label{fig:injectionlocking}
\end{figure}

Figure \ref{fig:injectionlocking}(a) shows the scheme used in injection locking a semiconductor laser to an external optical seed -- originating from a separate low noise laser. Although the utilized semiconductor laser comes with internal optics to improve beam symmetry, as is usual, the output beam shape still has a mild ellipticity.  The seeding laser beam, on the other hand, has a circular beam profile as it comes out of an optical fiber. In this demonstration, we use a 780-nm DFB laser to-be-injected (Toptica-Eagleyard, \SI{80}{\milli\watt}). The aspect ratio of its output is about 1.25. The seeding beam originates from a frequency doubled 1560-nm laser (OEwaves 4030 series) coupled into a polarization maintaining fiber, providing \SI{230}{\micro\watt} of seed power at 780 nm. About 6\% of the laser output is picked off using a beam sampler, and used for injection status monitoring with a  QPD (FirstSensor QP50-6-SM). The QPD is mounted on a miniature 2-axis translation stage allowing precise centering of the QPD on the beam. Reading out the diagonal channel of the QPD (Fig. \ref{fig:injectionlocking}(b)) gives a measure of the beam's ellipticity.  


The QPD is operated using a home-built transimpedance amplifier circuit. If the QPD diagonal channel were to be implemented in  the `balanced' operation $(A+B)-(C+D)$, with A, B, C, D being the optical powers measured at the four quadrants, a circularly shaped incident beam would have resulted in a zero signal -- as in the operation described in \cite{SquashLockArxiv}. The signal would change to either positive or negative values as the beam ellipticity is varied. In contrast, the diagonal channel in the new `asymmetric' operation utilized in the current work is weighted as $\varepsilon_1(A+B)-\varepsilon_2(C+D)$. The individually adjustable weights $\varepsilon_1$ and $\varepsilon_2$ allow us to electronically change the effective ellipticity read-out, such that we can get a zero signal for the slightly elliptical beam shape of our laser's output in absence of optical seeding.

\begin{figure}[!tb]
    \centering\includegraphics[]{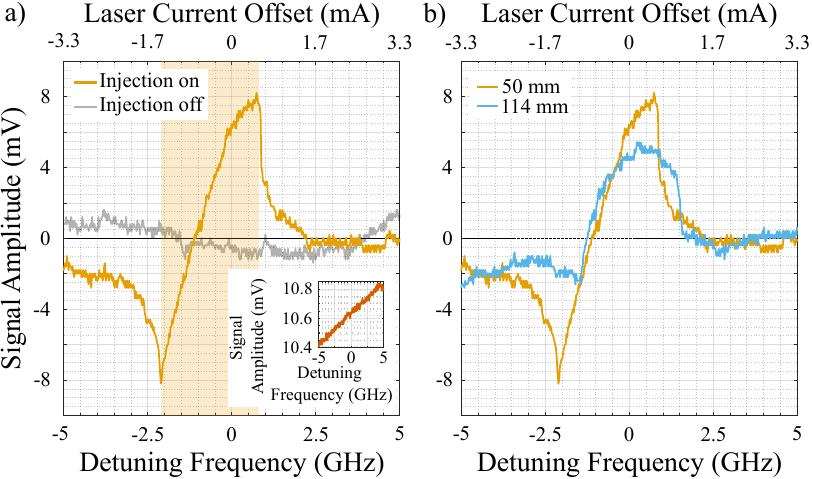}
    \caption{a) A typical error signal as a function of the frequency detuning between the free running laser and the seeding light frequency -- negative numbers indicating lower seeding laser frequency. `Injection off' data was taken with the seeding laser beam blocked. The frequency span under the error signal (shaded region) corresponds to the optimally injection locked range. The inset shows the output of the sum channel $(A+B+C+D)$ of the QPD. b) The error signal at two selected distances between the QPD and the beam sampler, illustrating the Gouy phase shifts acquired in the setup.}
    \label{fig:results_data}
\end{figure}

\begin{figure}[!t]
    \centering\includegraphics[]{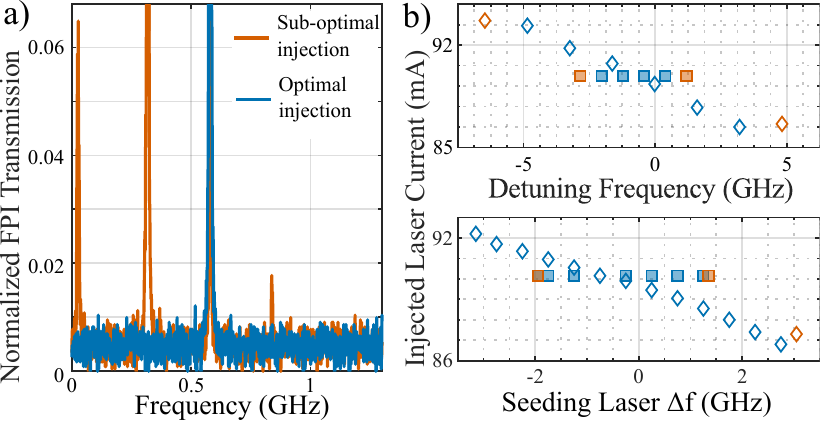}
    \caption{a) Example optical spectra observed from the FPI (1.5 GHz free-spectral-range) showing optimal (blue) and sub-optimal (orange, offset for clarity) laser injection locking. Traces are normalized to set the injected signal peak to 1. b) Monitored laser current and injection locking status as a function of injected laser temperature {expressed in units of absolute detuning frequency (top)} and seeding laser frequency (bottom) -- with (diamonds) and without (squares) engaging feedback. Blue: optimally injected, Orange: sub-optimally injected. Temperature is converted to frequency units as per the temperature dependence of the free-running laser frequency. Scanned ranges are $\sim$0.2$^\circ C$ and $\sim$0.1$^\circ C$ for the cases with and without the feedback engaged.}
    \label{fig:lock_data}
\end{figure}

In order to demonstrate an error signal for injection locking status, the laser current was swept using a triangular wave form. The laser temperature was set for the current scan to lay within a mode-hop-free tuning range of the laser. A typical signal is shown in figure \ref{fig:results_data}(a). The data shown is the output of the diagonal channel passed through a 4 kHz lowpass filter with the QPD operated in the asymmetric configuration. In absence of the injection beam, no significant change in the beam ellipticity is observed. However, upon injection, the change in ellipticity from the spatial mode interference of the seeding and the laser beams leads to a variation in the diagonal channel output, forming an error signal that provides information about which side of the good injection band we are situated. The laser current band centered around the error signal (Fig. \ref{fig:results_data}(a)) was indeed verified to be the good injection locking range as observed from independent measurements of the laser spectrum with an FPI. In this laser current range, spurious laser tones and excess noise tails are absent, and the frequency of the seeding beam is tracked by the laser( Fig. \ref{fig:lock_data}(a))
. 


In obtaining the results presented in Fig. \ref{fig:results_data}(a), the alignment of the QPD and the asymmetric operation are crucial in obtaining a robust error signal.
In order to observe the correct error signal, it is important to first center the QPD on the beam. The relative position can be monitored using the left-right (LR: $(A+C)-(B+D)$ ) and the up-down (UD: $(A+D)-(B+C)$) channels of the QPD. In principle, zeroing out these signals should decouple the diagonal channel from any fluctuations in the beam position. In practice however, we need to further fine-tune the QPD position such that the crosstalk of small misalignments into the diagonal channel output is suppressed -- i.e. where the diagonal channel is at a quadratic turning point as a function of QPD translations. Such discrepancies in the UD and LR channel outputs are likely due to small electronic asymmetries or non-uniform photo-current conversion between different quadrants of the QPD. This alignment procedure ensures insensitivity to small fluctuations in beam position. 

If the diagonal channel of the QPD were to be operated in the balanced configuration, majority of the background ellipticity would come from the elliptical beam shape of the laser. In comparison, the relevant change in ellipticity due to the spatial interference between the laser mode and the seeding mode would be very small. Importantly, in this configuration, the background ellipticity signal would lead to a direct bleeding of intensity fluctuations into the diagonal signal. However, in the asymmetric QPD operation, where there is no background signal, laser intensity fluctuations decouple from the diagonal signal. We note that the asymmetric operation brings an undesired sensitivity to ambient light changes, nevertheless a cover on the QPD, blocking ambient light, eliminates this problem. We further note that the background ellipticity can also be eliminated by tilting the QPD such that the elliptical beam is projected as a circular beam on the QPD. This approach is insensitive to ambient lighting conditions, but would require a fine control of the QPD orientation. For the results shown, only a small tilt is present to prevent any back reflections for robust operation.

Insensitivity to laser power fluctuations for a properly tuned setup is illustrated in the inset of figure \ref{fig:results_data}(a). The laser power was monitored using the output of the sum channel on the QPD, which gives the total power of the four quadrants, i.e., $A+B+C+D$. Although the sum channel shows about 4\%  change over the current scanning range, the error signal on the diagonal channel output is insensitive to these power changes. This is especially evident when observing the `injection off' curve (gray curve in figure \ref{fig:results_data}(a)) which remains unaffected by the power change. In comparison, for the symmetric QPD operation, the crosstalk with the intensity would have dominated the signal magnitude. Also note that no significant sensitivity to beam pointing changes due to air currents or thermal drifts was observed after proper alignment.


For efficient operation of the setup, the injected beam mode matching needs to be optimized. This optimization is done in two steps. We first ensure that the seeding laser beam is travelling along the same optical axis as the laser beam after passing through the optical isolator. To achieve this, the laser output is first coupled into an optical fiber (Fig. \ref{fig:injectionlocking}(a)), and then the seeding laser beam is aligned such that it optimally couples to the same fiber after reflecting from the laser cavity -- but with the laser power switched off. The laser power is then switched back on to operate the system. 

The shape of the obtained error signal is in fact a function of the distance from the laser to the QPD due to the different Gouy phase shifts acquired by the laser mode and the 2$^{nd}$ order reference modes that interfere to form the error signal. Figure \ref{fig:results_data}(b) shows the Gouy phase shifts observed in this setup by varying the QPD-to-beam sampler distance. The differential Gouy phase shift changes by approximately $\pi/2$ between the two shown curves, going from a near-dispersive (orange curve) to a near-absorptive (blue curve) signal. The near-dispersive signal is the useful one for generating a feedback on the laser current for injection lock stabilization, indicating that there is a preferred range of distances from the laser to the QPD for the developed injection locking monitoring scheme. This range experimentally turned out to be in the 10-20 cm scale; rather practical. In principle, any error signal shape with a zero-crossing within the injection range can be used for stabilizing the injection current.

The observed Gouy phase shifts are in fact taking place at a distance scale an order of magnitude smaller than those expected given the collimation properties of the laser output and the injection beams. Currently, a modeling of this unexpected behaviour is not feasible without a detailed knowledge of what exact collimation optics is integrated into the proprietary packaging of the utilized laser module. 


Having discussed the details and the practical considerations for error signal generation, we now turn to the stabilization of injection locking. To this end, a home-built kilohertz-bandwidth proportional-integral controller (PI in Fig. \ref{fig:injectionlocking}(a)) is utilized to feedback onto the laser current to lock the error signal at its zero crossing. The resulting stabilization keeps the laser current within the optimal range for injection locking against varying external factors, and extends the injection locking range. To demonstrate the robustness brought by the stabilization technique, we characterize the operation in two ways: we vary the temperature of the injected laser, or we vary the seeding laser frequency. At all times we monitor the laser current, and independently the optical spectrum as observed by the FPI. This allows us to characterize the increase in the parameter ranges in which the optimal injection persists when the feedback is engaged (Fig. \ref{fig:lock_data}(b)). For example, in the case of temperature scanning, we observe an increase in the range for optimal injection from $\sim$3 GHz to $\sim$9 GHz (equivalent to 80\% of the mode hop-free current tuning range of the free-running laser). We also see that this is accomplished through the change in the laser current induced by the feedback. Here we define `optimal injection locking' practically through the lack of any measurable spurious tones in the optical spectrum. Representative examples of optimal and sub-optimal cases are illustrated in Fig. \ref{fig:lock_data}(a) and color coded as blue and orange cases respectively throughout the figure. Spurious tones appear in the sub-optimal case in addition to the single tone which tracks the seed frequency in the optimal case.

In this letter, we unveiled a powerful method for active stabilization of laser injection locking, based on the recently developed 'squash locking' technique \cite{SquashLockArxiv}. The method utilizes beam shape mismatch between the seeding beam and the seeded laser. Due to interference effects, this mode mismatch leads to a varying output beam ellipticity across the injection locking parameter range as the laser parameters or injection frequency are varied. This is then utilized to feedback to laser properties to keep the laser parameters at the optimal range required for successful injection locking. Compared to existing techniques for injection locking monitoring, the method presented here requires no internal additions or modifications to an existing injection locked laser setup. The method is modulation free, plug-and-play for existing systems, compact, and insensitive to power and beam pointing fluctuations.

Although not discussed here, similar error signals can also be obtained by scanning the temperature of the injected laser at a constant current. This suggests that the feedback can also be done on the laser temperature instead of the current in order to further extend the optimal injection locking range. Conceptually, it is possible to extend the developed method in a straightforward manner to self-injection-style laser systems such as ECLs formed by gratings, resonators, and other frequency selective self-feedback elements. For example, in grating-style ECLs where the grating angle defines the spectral filtering properties \cite{Schremer1990}, a similar error signal can be obtained to feedback on the grating angle instead of the laser current. It is also possible to apply this method to other laser types besides semiconductor lasers. The only requirement is that there is a mode shape mismatch between the laser output mode and the (self-)seeding mode. The method laid down in this work paves the way towards robust laser systems that avoid mode-hops.

\begin{backmatter}
\bmsection{Funding} This work was supported by Institute of Science and Technology Austria.

\bmsection{Acknowledgments} OH and FD conceived the experiments. UM performed the experiments. UM, FD and OH analyzed the data and prepared the manuscript. VL built the initial injection locked laser system -- without injection locking monitoring. SW, SA and FD provided experimental support and helped UM design the optical setup and the electronics feedback circuits.

\bmsection{Disclosures} OH, FD: Institute of Science and Technology Austria (P).

\bmsection{Data availability} Data underlying the results presented in this paper are not publicly available at this time but may be obtained from the authors upon reasonable request.


\end{backmatter}

\bibliography{sample}


\end{document}